\documentclass[journal]{IEEEtran}

\usepackage{cite}
\usepackage{amsmath,amssymb,amsfonts}
\usepackage{algorithmic}
\usepackage{graphicx}
\graphicspath{ {figures/} }
\usepackage{textcomp}
\usepackage{xcolor}
\usepackage[bookmarks=false]{hyperref} 
\usepackage{algorithm}
\usepackage{booktabs}
\usepackage{multirow}
\usepackage[separate-uncertainty,tight-spacing=true]{siunitx}
\usepackage{blindtext}
\usepackage{subcaption}
\usepackage{multicol}
\usepackage{tabularx}
\usepackage{caption}
\usepackage{array}
\newcolumntype{P}[1]{>{\raggedright\arraybackslash}p{#1}}

\usepackage{lipsum}

\newcolumntype{Y}{>{\centering\arraybackslash}X}

\begin{document}

\title{Trust-based Blockchain Authorization for IoT}

\author{Guntur~Dharma~Putra,~\IEEEmembership{Student~Member,~IEEE,}
        Volkan~Dedeoglu,
        Salil~S~Kanhere,~\IEEEmembership{Senior~Member,~IEEE,}
        Raja~Jurdak,~\IEEEmembership{Senior~Member,~IEEE,}
        and~Aleksandar~Ignjatovic

\thanks{G.D. Putra, S.S. Kanhere, and A. Ignjatovic are with the School of Computer Science and Engineering, the University of New South Wales Sydney (UNSW), Australia and also with Cyber Security Cooperative Research Centre (CSCRC), Australia (e-mail: {gdputra, salil.kanhere}@unsw.edu.au, ignjat@cse.unsw.edu.au).}
\thanks{V. Dedeoglu is with the Commonwealth Scientific and Industrial Research Organisation's Data61, Pullenvale, Australia (e-mail: volkan.dedeoglu@data61.csiro.au).}
\thanks{R. Jurdak is with the School of Computer Science, Queensland University of Technology (QUT), Brisbane, Australia (e-mail: r.jurdak@qut.edu.au).}
}


\maketitle

\begin{abstract}
Authorization or access control limits the actions a user may perform on a computer system, based on predetermined access control policies, thus preventing access by illegitimate actors. Access control for the Internet of Things (IoT) should be tailored to take inherent IoT network scale and device resource constraints into consideration. However, common authorization systems in IoT employ conventional schemes, which suffer from overheads and centralization.
Recent research trends suggest that blockchain has the potential to tackle the issues of access control in IoT. However, proposed solutions overlook the importance of building dynamic and flexible access control mechanisms.
In this paper, we design a decentralized attribute-based access control mechanism with an auxiliary Trust and Reputation System (TRS) for IoT authorization. Our system progressively quantifies the trust and reputation scores of each node in the network and incorporates the scores into the access control mechanism to achieve dynamic and flexible access control. We design our system to run on a public blockchain, but we separate the storage of sensitive information, such as user's attributes, to private sidechains for privacy preservation. We implement our solution in a public Rinkeby Ethereum test-network interconnected with a lab-scale testbed. Our evaluations consider various performance metrics to highlight the applicability of our solution for IoT contexts.
\end{abstract}

\begin{IEEEkeywords}
trust management, privacy, authorization, blockchain, IoT, reputation.
\end{IEEEkeywords}

\IEEEpeerreviewmaketitle

\section{Introduction}
\label{sec:intro}
\IEEEPARstart{P}{rotecting} important resources from illegitimate access has been one of the priorities in securing computer systems. The protection mechanism, which is commonly referred to as access control or authorization, decides when to grant or deny an access request from an authenticated user based on certain access policies, which determine what operations the user is permitted to perform~\cite{sandhu1994access}. There are various models of conventional access control, for instance, Role-Based Access Control (RBAC), Attribute-Based Access Control (ABAC), and Capability-Based Access Control (CapBAC)~\cite{Ouaddah2017}, where in general access levels are mapped to certain metrics, namely user's roles, attributes, and capabilities (access tokens) respectively. While these conventional notions are also applicable to Interent of Things (IoT) ecosystems, incorporating the same in practical deployments raises concerns due to the inherent constraints of IoT devices such as limited power resources, memory size, and computational capacity.

Recent research has shown that blockchain has the potential to overcome a number of unresolved problems related to access control in IoT, such as single point of failure~\cite{Novo2018}. For instance, ABAC mechanisms were implemented in a decentralized fashion to provide more fine grained access control by recording attribute registrations and revocations in blockchain transactions~\cite{Ding2019}. In this scheme, access control is enforced by the resource owner by searching the blockchain for such records.
In addition, the authors in~\cite{Jiang2018} proposed a transparent access control mechanism wherein a list of static access rights are stored in the blockchain and enforced by a smart contract.

While these proposals have addressed the issues in authorization systems, such as a single point of failure and lack of transparency, such static authorization schemes are unable to automatically capture the dynamics of the network and adapt their authorization policies accordingly. In the case of unwanted circumstances in the network, such as a node being compromised, static authorization schemes would continue to enforce predefined access control policies that assume normal behavior rather than making proper adjustments to account for compromised conditions. The inability of these static authorization schemes to reinforce adjustments in the access control policies, in fact, raises a critical question of how to achieve a dynamic and trustworthy access control system without overlooking the fact that access control requires a sensitive consideration of who can access a resource.

In this paper, we propose a dynamic authorization scheme by designing a decentralized ABAC system enriched with a novel Trust and Reputation System (TRS). Our approach quantifies both Service Consumer (SC) and Service Provider (SP) behavior while also simplifying the trust and reputation score computation using recursion. In our TRS, the trustworthiness and reputation of a node are calculated based on their adherence to the access control policies. We define trust as a subjective belief of a node's behavior based on the previous interactions towards another node, which may help to determine the likelihood of the next interaction. On the other hand, we refer to reputation as a global view of a node's past behavior from aggregated trust relationships from multiple nodes~\cite{DiPietro:2018:BTS:3205977.3205993}. In TRS, both trust and reputation scores are transparently calculated by smart contracts in the main public blockchain. It is important to note that progressive evaluation of trust and reputation scores may help to detect and eliminate malicious or compromised nodes in the network~\cite{MohamadNoor2019}.
In addition, we propose a clear separation of sensitive information storage, such as users' attributes. We design a multi-tier decentralized architecture, which consists of a main public blockchain to enforce access control via smart contracts, and additional private sidechains to store sensitive information securely. We implement a proof-of-concept of our proposed solution and test it on a Rinkeby Ethereum test-network interconnected with a local IoT test-bed comprised of Raspberry Pi devices. However, our architecture design is agnostic to the blockchain platform with the only requirement being that it should support smart contracts.

In summary, the contributions of the paper are as follows:
\begin{itemize}
    \item We design a decentralized IoT access control mechanism, which is based on ABAC with an auxiliary TRS, dubbed as blockchain-based trust management (BC-TRS), to capture the dynamics of the network. Our architecture separates the storage of sensitive information, e.g., user's attributes, and the public TRS data, by incorporating a main public blockchain and additional private sidechains, and supports asynchronous authorization.
    \item We design a lightweight TRS which involves a simple recursive calculation that captures  bi-directional interactions of SC and SP.
    \item We develop a proof-of-concept implementation of our proposed solution in a public Rinkeby Ethereum test-network interlinked to a lab-scale IoT test-bed. To demonstrate the practicability of our proposed solution in the IoT context, we evaluate our solution in terms of trust and reputation score evolution, authorization latency, and Ethereum gas consumption.
\end{itemize}

The rest of the paper is organized as follows. Section~\ref{sec:related-works} presents related works in blockchain and IoT access control. We describe our proposed system model, TRS, and access control mechanism in Section~\ref{sec:system-model},~\ref{sec:trust-and-reputation},~and~\ref{sec:ac-framework}, respectively. We present our proof-of-concept implementation and the corresponding results in Section~\ref{sec:performance-evaluation}. We discuss our findings in Section~\ref{sec:discussion} and conclude the paper in Section~\ref{sec:conclusion}.


\section{Related Works}
\label{sec:related-works}
Recent research has shown that decentralization in IoT access control may overcome some limitations of conventional access control mechanism such as reliance on trusted third parties and centralized processing.
In~\cite{Andersen2019}, Andersen et. al presented a scalable decentralized authorization scheme, wherein access right is given to the requester via cryptographic access delegation method. A reverse-discoverable decryption mechanism is implemented to ensure the privacy of the access, which may span across different administrative domains.
In~\cite{Pinno2018}, the authors proposed a full decentralized IoT access control system comprised of four interconnected blockchains, namely, accountability, context, relationships, and rules blockchains. The framework stores entities and authorization information as blockchain transactions and supports three types of access control mechanisms, namely Access Control List (ACL), CapBAC, and ABAC.
In~\cite{DORRI2019180}, a Lightweight Scalable blockchain (LSB) is proposed to deliver end-to-end security and access management for IoT via an ACL. LSB consists of interconnected and independent clusters, wherein a cluster head stores and maintain an ACL, using which access request is validated.
Ding et. al.~\cite{Ding2019} proposed a framework for decentralized ABAC, where blockchain transactions play a significant role in the authorization and revocation of attributes. In their framework, the resource owner is responsible for enforcing the access control policies, wherein, the attribute validation process is performed by searching the blockchain. However, these proposals~\cite{Pinno2018, Andersen2019, DORRI2019180, Ding2019} overlook the full potential of blockchain, e.g., leveraging the capabilities of smart contracts, and mainly rely on off-chain processes.

To fully utilize blockchain's potential in IoT access control, some proposals have designed frameworks that aim to deliver decentralization by means of smart contracts. Zhang et. al.~\cite{Jiang2018} proposed a framework that uses smart contracts to replace the centralized validation of access policies. The mechanism consists of three smart contracts, namely access control contracts, a judge contract, and a register contract, in which access control policies are stored as ACLs.
In~\cite{Andersen2017}, the authors proposed a new concept, called Delegation of Trust (DoT), which embodies the notion of trust that a resource owner has in a legitimate user to access the owner's resource.
In~\cite{Novo2018}, Novo designed an architecture that utilizes a smart contract but only for managing access control in a permissioned blockchain. The design requires IoT device managers to execute a function call in the smart contract to authorize service consumers by checking if the access request aligns with the ACLs.
Pal et. al. proposed a blockchain-based access control mechanism for IoT with built-in support of access delegation enforced by smart contracts~\cite{8894097}. The mechanism leverages identity-less and asynchronous authorization, in which access request is validated against privately stored attributes that preserve user's privacy.
While these proposals~\cite{Jiang2018, Novo2018, Andersen2017, 8894097} use smart contracts to achieve decentralization of access control logic, they are unable to  capture the network dynamics that are inherent in IoT ecosystems.

A trust computation model may protect against unfair or malicious activities from misbehaving nodes in a network~\cite{Zhang2018}. To date, there have been some trust management protocols for IoT~\cite{Chen2016, Chen2016-soa, BernalBernabe2016, Gwak2018, icbc2020, Chen2019, Al-Hamadi2019, Kouicem2019, Daoud2019, 10.1145/3360774.3360822}, some of which are mainly tailored for inclusion in IoT access control mechanisms. In~\cite{BernalBernabe2016}, the authors proposed a trust computation model for a trust-aware access control mechanism for IoT.
Gwak et. al. proposed TARAS, an adaptive role-based access control for IoT that incorporates trust computation for granting or denying access without requiring any prior knowledge of the requester~\cite{Gwak2018}.
In~\cite{icbc2020}, the authors designed a decentralized trust-aware access control mechanism for IoT, where a customized ABAC is employed with an additional trust management systems to quantitatively assess the trustworthiness of the requester according to the prior experience.
Di Pietro et. al. proposed an access control mechanism where both requester and requestee should agree upon particular terms and obligations on resource access, which also incorporate the requestee's global reputation~\cite{DiPietro:2018:BTS:3205977.3205993}.
However, the model~\cite{DiPietro:2018:BTS:3205977.3205993} is relatively complex and inefficient, which is unsuited for IoT. Some proposals~\cite{icbc2020, Gwak2018} also disregard the importance of privacy preservation and centralized Trusted Third Party (TTP) processing is still used~\cite{BernalBernabe2016, Gwak2018}.

\begin{table*}
\centering
\caption{Overview of related works.}
\label{ta:comparison-blockchain-auth}
\begin{tabular}{lcccccc}
\toprule
\multirow{2}{*}{\textbf{Proposal}} & \textbf{Smart} & \textbf{Privacy} & \textbf{Trust} & \textbf{Asynchronous} & \textbf{Authorization} \\

 & \textbf{Contracts}& \textbf{Preservation} & \textbf{Computation} & \textbf{Authorization} & \textbf{Type} \\
\midrule
Novo~\cite{Novo2018} & For access management & N/A & N/A & N/A & ACL \\

Ding~\cite{Ding2019} & N/A & N/A & N/A & N/A & ABAC \\

Zhang~\cite{Jiang2018} & To implement ACL & N/A & N/A & N/A & ACL \\ 

Di Pietro~\cite{DiPietro:2018:BTS:3205977.3205993} & N/A & Encryption & N/A & For reputation update & Terms and obligation \\

WAVE~\cite{Andersen2019} & N/A & Encryption & N/A & N/A & Authorization graph \\

ControlChain~\cite{Pinno2018} & N/A & N/A & N/A & N/A & ACL, ABAC, CapBAC \\

LSB~\cite{DORRI2019180} & N/A & Changeable PK & N/A & N/A & ACL \\

WAVE~\cite{Andersen2017} & For authorization layer & Encryption & N/A & For data retrieval & Permission graph \\

Pal~\cite{8894097} & For handling delegation & Sidechains & N/A & Supported & ABAC \\

TACIoT~\cite{BernalBernabe2016} & N/A & Security Policy & Supported & N/A & XACML \\

TARAS~\cite{Gwak2018} & N/A & N/A & Supported & N/A & RBAC \\

IoT TM~\cite{icbc2020} & For authorization layer & N/A & Supported & N/A & ABAC \\
    
BC-TRS & For authorization and trust mgmt. & Sidechains & Supported & Supported & ABAC \\

\bottomrule
\end{tabular}
\end{table*}

We present an overview of related works in Table~\ref{ta:comparison-blockchain-auth}. In conclusion, previous studies overlook the importance of capturing the dynamic interactions of IoT nodes in access control mechanism, i.e., via the means of trust computation.

\section{System Model}
\label{sec:system-model}
In this section, we illustrate the proposed decentralized trust management for IoT access control and describe the main components and threat model of our proposed system in detail.

\begin{figure*}
\centering
\includegraphics[width=0.8\textwidth]{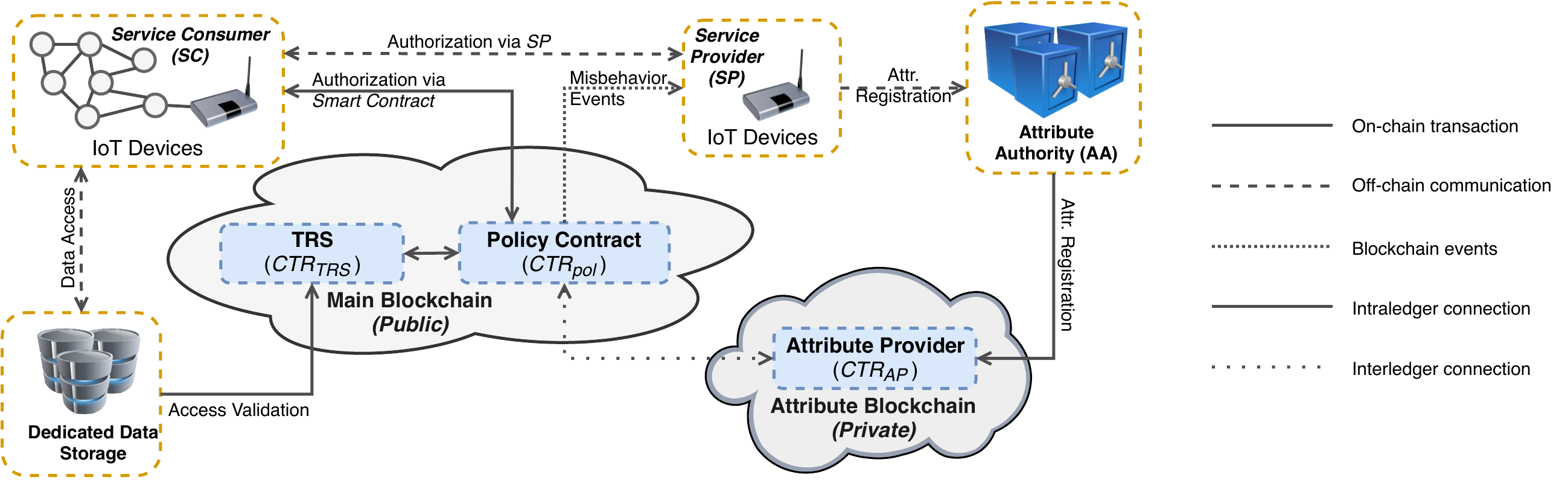}
\caption{Architecture overview.}
\label{fig:architecture-overview}
\end{figure*}

\subsection{Main Components}
\label{subsec:main-components}
    We consider a network model, wherein each IoT device is registered to a local regulator that maintains a list of technical specifications and ownership information of the corresponding device (i.e., device attributes). To manage and limit who can access which resources under certain conditions, an attribute-based access control mechanism is employed. Fig.~\ref{fig:architecture-overview} illustrates our proposed trust management and decentralized access control architecture. To support separation of sensitive information, we utilize two types of blockchain networks, a public and a private blockchain, in which different categories of data are stored.
    
    \subsubsection{Service Providers and Service Consumers}
    \label{subsubsec:sp-and-sc}
    In general, the IoT nodes are grouped into two categories, namely $SPs$ and $SCs$. Typically, $SP$s own a set of resources, denoted $r$, which can be accessed by others for free or by paying for a small access fee. Meanwhile, $SC$s are user devices interested to consume resources $r$ owned by $SP$s. For identification purposes, we use Public Keys ($PK$s) for both $SP$s and $SC$s.
    The IoT devices run a light blockchain client to directly connect to the public blockchain network~\cite{elktechcrunch}, which implies that the devices should have sufficient resources to support asymmetric cryptography. In instances where devices are resource constrained, they may opt to use a third party service or rely on their communication gateway for providing connectivity to the blockchain. 
    
    \subsubsection{Dedicated Data storage}
    \label{subsubsec:dds}
    The off-chain Dedicated Data Storage ($DDS$) stores high volume of data over a longer timespan. The $SC$s store and update the data in a regular basis with an additional signature of $SC$ to ensure integrity. We assume that there are multiple $DDS$ in the network with sufficient redundancy level to provide high availability and maintain scalability.
    An $SC$ may request access to the main blockchain for obtaining a legitimate access token for accessing the data in the $DDS$. We describe the access control framework in more detail in Section~\ref{sec:ac-framework}.
    
    \subsubsection{Attribute Authorities}
    \label{subsubsec:attribute-authority}
    We use inherent attributes of an IoT device (e.g., sensor types and hardware specifications) to determine whether a node is allowed to access a resource, as defined in an ABAC scheme.
    The Attribute Authority ($AA$) is responsible to issue legitimate attributes to the participating IoT nodes, according to the prescribed guidelines that conform to hardware specifications and ownership information. There are multiple $AA$s that form decentralized attribute authority networks, using which an $SC$ can request for attributes issuance. We describe the attribute registration process in detail in Section~\ref{subsec:ac-primitives}.
    
    \subsubsection{Blockchain Networks}
    \label{subsubsec:blockchain-network}
    In our proposed model, we implement a single main public blockchain, denoted $MB$, which provides decentralized and collaborative trusted execution of access control logic and tamper-proof trusted data storage. 
    As publicly revealing IoT attributes may pose some privacy concerns, we implement a set of permissioned blockchains, $PB = \{pb_1, pb_2, pb_3, \ldots, pb_N\}$, to maintain a private immutable list of attribute records. While $MB$ is maintained by independent miners that are motivated to gain financial benefits from mining the blocks, each $pb_k$ is maintained by a consortium of independent and partially-trusted attribute authorities $AA_k = \{AA_k^1, AA_k^2, \ldots, AA_k^Y\}$.
    Note that, access to $pb$ is limited to the corresponding $AA$s. In addition, each $pb$ is interlinked to the main blockchain, hence, acting as a sidechain for supplying attribute information to the main blockchain.
    
    Note that $MB$ can also be implemented as a permissioned blockchain network, in which the participants may be known in advance and may be partially trusted. However, a more stringent access restriction should apply to $PB$ to avoid leakage of sensitive information, i.e., no $MB$ participants are allowed to access $PB$ as it should only be accessible by the corresponding $AA$.
    
    We deploy two public smart contracts to $MB$, namely TRS ($CTR_{TRS}$) and Policy Contract ($CTR_{pol}$), which store the logic of trust calculation and access policy validation, respectively. In addition, a private smart contract, namely Attribute Provider Contract ($CTR_{AP}$), which resides in each $pb$, is responsible for attribute registration and validation. As such, we have $N$ number of $CTR_{AP}$, i.e., $\{CTR^1_{AP}, CTR^2_{AP}, \ldots, CTR^N_{AP}\}$. We employ a bridging mechanism to interlink $MB$ and $PB$, with the assumption that both types of blockchains are identical in terms of technical implementation~\cite{Siris2019}, e.g., use the same public key cryptography mechanism to handle authentication and signature process. The bridging mechanism works by matching the $PK$s from both blockchains.

\subsection{Threat Model and Assumptions}
\label{subsec:threat-model}
    In our architecture, the adversaries can be $SP$s or $SC$s that are maliciously intent on disrupting the network. We group our threat model into three categories, namely access control policies attacks, reputation attacks, and other network protocol attacks. Firstly, a malicious $SC$ may perform attacks related to access control, for instance, performing a replay attack in which the adversary maliciously captures and reuses an access token to gain illegitimate access. A malicious $SC$ may also successively try to access the data using forged and expired access tokens or to leverage access rights, which would result in network congestion, i.e., a DoS attack. In addition, a malicious $SP$ may be unreliable in providing the data, e.g., an $SP$ does not fulfil its obligation in providing frequent updates of sensor readings. 
    Secondly, the adversaries are also capable of performing the following types of reputation attacks:

    \begin{itemize}
        \item \textit{Self-promoting attacks:} As an IoT node may act as both $SP$ and $SC$, a malicious actor might try to increase its own reputation score by providing positive feedback to itself.
        \item \textit{Bad-mouthing attacks:} A malicious $SC_i$ may attempt to ruin the reputation of an $SP_j$ by constantly providing negative feedback regardless of the quality of the service. Moreover, an $SC_i$ may also attempt to ruin the trust and reputation of an $SC_k$ by requesting illegitimate access on behalf of $SC_k$.
        \item \textit{Ballot-stuffing attacks:} A malicious $SC_i$ may collude with $SP_j$ to increase their reputation scores. 
        \item \textit{Whitewashing or newcomer attacks:} An $SC$ attempts to rejoin the network using a new identity aiming to maliciously erase its previously recorded bad behavior and obtain a fresh reputation score.
    \end{itemize}
    
    Third, we presume other types of network protocol attacks may emerge. However, those attacks are handled by established intrusion detection mechanisms and thus beyond the scope of this paper, while the other two groups of attacks are handled by our TRS mechanism.
    
    As our system model primarily runs on top of a commodity blockchain platform, we assume that the blockchain is secure against peer-to-peer and consensus attacks, such as eclipse, Sybil, and 51\% attacks~\cite{8543246}.
    We further presume that $AA$s are partially trusted and are not required to be always connected to the system. However, to maintain the security we require $3f+1$ online $AA$s at any given time, where $f$ is the number of faulty or untrusted $AA$s in $pb$. 
    We also assume that $DDS$ is secure and allows only authorized users to access the stored data.

\section{Trust and Reputation System}
\label{sec:trust-and-reputation}
We design the proposed TRS to achieve a dynamic and self-adaptive authorization system which is capable of capturing the dynamics of the network and detecting and eliminating malicious or compromised nodes,
as explained in our threat model (Section~\ref{subsec:threat-model}). If a malicious activity occurs, either $CTR_{pol}$ or $CTR_{TRS}$ will blacklist the offending node from the network and notify all network participants via blockchain events.

\begin{figure}
\centering
\includegraphics[width=0.3\textwidth]{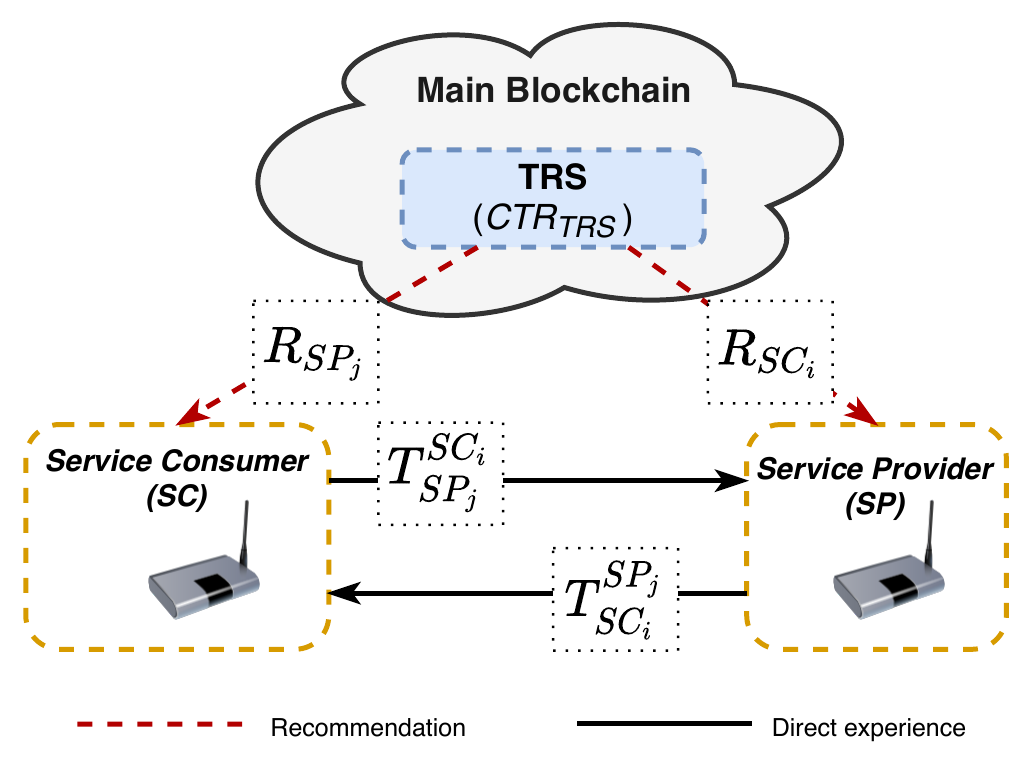}
\caption{The trust relationship model.}
\label{fig:trust-model}
\end{figure}

Fig.~\ref{fig:trust-model} illustrates the trust relationship model between the $CTR_{TRS}$ and IoT nodes ($SP$ and $SC$). The trust score of $SP_j$ towards $SC_i$, denoted $T^{SP_j}_{SC_i}$, is calculated by $CTR_{TRS}$ based on their previous interaction, which corresponds to a binary experience, i.e., either positive or negative. A positive experience is an honest action of $SC_i$ to $SP_j$ that conforms to the access control policies, while negative experience is otherwise. An initial value of $0$ is assigned to $T^{SP_j}_{SC_i}$, if there are no prior interactions of $SC_i \rightarrow SP_j$. We calculate $T^{SP_j}_{SC_i}$ after $t$  interactions as follows:
\begin{equation}
\label{eq:trust-formula-sc}
    T^{SP_j}_{SC_i}(t)=(1-\gamma)\sum_{m=1}^t \delta_m\gamma^{t-m}    
\end{equation}
where
\begin{equation*}
    \delta_m=\begin{cases} \delta_{pos}& \text{if \ } m^{th} \text{ \ interaction was positive}\\ \delta_{neg}&\text{otherwise}\end{cases}    
\end{equation*}
where $\gamma$ is the ageing parameter ($0< \gamma < 1$) which affords more weight to recent observations than older ones. $\delta_{pos}>0$ is the weight associated with a positive interaction, while $\delta_{neg}<0$ represents the weight for negative interactions. We choose  $\delta_{pos}<|\delta_{neg}|$, to make it harder to build trust than to lose it, as how humans perceive trust in real-life social relationships.

Note that in the extreme case where all interactions are positive, we have:
\begin{equation}
\label{eq:extreme-cases}
    T^{SP_j}_{SC_i}(t) = (1-\gamma)\delta_{pos}\sum_{k=0}^{t-1}\gamma^k
\end{equation}
As $t \rightarrow \infty$ and all interactions are positive, we have the limiting case:
\begin{align}
    T^{SP_j}_{SC_i}(\infty) &= (1-\gamma)\delta_{pos}\sum_{k=0}^{\infty}\gamma^k \nonumber \\
    &= \delta_{pos}
\end{align}
Similarly, in the other extreme case where all experiences are negative, we obtain $T^{SP_j}_{SC_i}(\infty) = \delta_{neg}$. The trust scores are therefore bounded by these two limiting cases:  
\begin{equation}
\label{eq:trust-boundaries}
    \delta_{neg} \leq T^{SP_j}_{SC_i}(t) \leq \delta_{pos}
\end{equation}
where $T^{SP_j}_{SC_i} = \delta_{neg}$ is the score when $SC_i$ is totally untrusted and  $T^{SP_j}_{SC_i} = \delta_{pos}$ is the score when $SC_i$ is completely trusted by $SP_j$.

Besides being conveniently bounded, the trust score has another advantage, in that it is computable by simple recursion:
\begin{align}
    T^{SP_j}_{SC_i}(t+1) &= (1-\gamma)\sum_{m=1}^{t+1} \delta_m\gamma^{t+1-m} \nonumber\\ 
        &=\gamma (1-\gamma)\sum_{m=1}^{t} \delta_m\gamma^{t-m}+(1-\gamma)\delta_{t+1} \nonumber\\
        &=\gamma\, T^{SP_j}_{SC_i}(t)+(1-\gamma)\,\delta_{t+1}
\end{align}

We follow a similar approach as in Eq.~\eqref{eq:trust-formula-sc} to calculate the trust score of a particular $SP$ from an $SC$'s standpoint, but use different weighting that results in differences in the evolution of trust. The trust score of $SC_i$ towards $SP_j$, denoted $T^{SC_i}_{SP_j}$, is derived from accumulated feedback from $SC_i$ after receiving the service, i.e., accessing a resource owned by $SP_j$. Similarly, the feedback corresponds to a binary expression of either positive or negative experience. The trust score of $T^{SC_i}_{SP_j}$ after $t$ feedback instances is calculated as follows:

\begin{equation}
\label{eq:trust-formula-sp}
    T^{SC_i}_{SP_j}(t)=(1-\mu  )\sum_{n=1}^t \varepsilon_n\mu^{t-n}    
\end{equation}
where
\begin{equation*}
    \varepsilon_n=
        \begin{cases}
            \varepsilon_{pos}& \text{if \ } n^{th} \text{ \ feedback was positive}\\
            \varepsilon_{neg}&\text{otherwise}
        \end{cases}    
\end{equation*}
where $0< \mu < 1$, $\varepsilon_{pos}>0$, $\varepsilon_{neg}<0$, and $\varepsilon_{pos}<|\varepsilon_{neg}|$.

We adopt Gompertz function to model the reputation growth~\cite{HUANG2014130}. As in real-life social interaction, reputation increases gradually after successive positive interactions and drops significantly after a negative interaction. In general, the reputation $R_{SC_i}$ of $SC_i$ is calculated by feeding the aggregation of its trust scores $A_{SC_i}(t)$ across different $SP$s to the Gompertz function, as follows: 

\begin{equation}
    A_{SC_i}(t) = \frac{\ln |Peers_i(t)|}{|Peers_i(t)|}\sum_{SP_j \in Peers_i(t)}T^{SP_j}_{SC_i}(t)
\end{equation}
\begin{equation}
\label{eq:reputation-sc-gompertz}
    R_{SC_i}(t) = ae^{-be^{-cA_{SC_i}(t)}}
\end{equation}
where $Peers_i(t)$ denotes the set of $SP_j$'s that have interacted with $SC_i$ until time $t$ and $|Peers_i(t)|$ denotes the cardinality of $Peers_i(t)$, i.e., the number of unique peers.
Note that, $A_{SC_i}(t)$ is equal to $ln|Peers_i(t)|$ times the mean of
$T^{SP_j}_{SC_i}(t)$ of all peers of $SC_i$. In this way, a larger number of $Peers_i(t)$ increases $A_{SC_i}$ but in a tempered fashion, preventing an $SC_i$ to achieve a large value of $A_{SC_i}$ by having a large value of $T^{SP_j}_{SC_i}(t)$ but only by interacting with a small number of peers.
In the Gompertz function~\eqref{eq:reputation-sc-gompertz}, $a$, $b$, and $c$ are the asymptote, the displacement parameter along \texttt{x}-axis, and the growth rate, respectively. This way, interactions with larger number of peers reinforce the reputation, but with a more appropriate sub-linear growth rate. Note that, the Gompertz function also guarantees that $R_{SC_i}$ is bounded between $1$ and $0$, which corresponds to high and low reputation, respectively.

The reputation $R_{SP_j}$ of $SP_j$ is calculated in an identical fashion as follows:
\begin{equation}
    A_{SP_j}(t) = \frac{\ln |Peers_j(t)|}{|Peers_j(t)|}\sum_{SC_i \in Peers_j(t)} T^{SC_i}_{SP_j}(t)
\end{equation}
and
\begin{equation}
\label{eq:reputation-sp-gompertz}
    R_{SP_j}(t) = ae^{-be^{-cA_{SP_j}(t)}}
\end{equation}

TRS is administered by $CTR_{TRS}$ on $MB$ by updating the trust and reputation scores on certain events. First, $T^{SP_j}_{SC_i}$ and $R_{SC_i}$ are updated when $SC_i$ requests an authorization to access a resource of $SP_j$. $T^{SP_j}_{SC_i}$ and $R_{SC_i}$ are also updated if an access control violation happens. Second, $T^{SC_i}_{SP_j}$ and $R_{SP_j}$ are updated when $SC_i$ sends its feedback after receiving service from $SP_j$. We describe how the trust model works in action in Section~\ref{sec:ac-framework}.

\section{Access Control Framework}
\label{sec:ac-framework}
This section presents the proposed adaptive access control framework, which employs TRS to manage authorizations and data access.

\subsection{Access Control Primitives}
\label{subsec:ac-primitives}
    We employ an ABAC scheme, which incorporates a novel TRS. In general, registration of attributes is mandatory for any $SC$ prior to authorization. An attribute $\alpha_i^m$ of $SC_i$ is denoted as:
        
    \begin{equation}
        { \alpha }_{ i }^{ m }=\left< key, type, val \right> 
    \end{equation}
    
    \noindent where $key$, $type$, and $val$ correspond to the name, type, and value of $\alpha_i^m$, respectively. Note that, it is common for an $SC$ to have a set of attributes, denoted $A_i=\{\alpha_i^1, \alpha_i^2, \ldots, \alpha_i^M\}$. 
    An $SC$ registers itself by sending an attribute registration request via a secure channel to an $AA_k^y$ of $pb_k$, with which the $SC$ is associated. $AA_k^y$ should have some underlying evidence to validate and issue attributes, such as a smart building specification, and to prevent attributes forgery. Upon receiving a registration request, $AA_k^y$ verifies the request and stores the attributes to $pb_k$ by invoking a transaction:
    \begin{equation}
        { TX }_{ reg }=\left[ { A }_{ i }|{ Sig }_{ SC_i }| timestamp | {  Sig }_{ AA_k^y } \right] 
    \end{equation}
    where ${ Sig }_{ SC_i }$ and ${ Sig }_{ AA_k }$ correspond to the signatures of $SC$ and $AA_k$, respectively.
    As $AA$s are partially trusted, the other online $AA$s in $pb_k$ need to validate $TX_{reg}$ against the prescribed guidelines of attribute issuance, i.e., if the attributes actually match the underlying evidence. If $TX_{reg}$ is valid, $AA_k^x$ submits an endorsement message $E_i^x=\left< H(TX_{reg}), Sig_{AA_k^x}\right>$ to $pb_k$, which contains $TX_{reg}$ hash $H(TX_{reg})$ and $AA_k^x$ signature $Sig_{AA_k^x}$. Note that to withstand faulty or untrusted $AA$, we require $3f+1$ online $AA$s at any given time, where $f$ is the number of faulty or untrusted $AA$s in $pb$. Consequently, $TX_{reg}$ should obtain a minimum of $2f+1$ endorsements to be considered valid~\cite{castro1999practical}. Upon successful validation by other $AA$s, $CTR_{AP_k}$ issues a sealing transaction $TX_{seal}$ to $MB$ as a proof that the attributes have been successfully registered:
    \begin{equation}
        { TX }_{ seal }=\left[ H(A_i) | E_i \right] 
    \end{equation}
    where $H(A_i)$ and $E_i$ correspond to the hash of $A_i$ and the collection of endorsements from other $AA$s, respectively.
    
    To explicitly define the requirements to access resource $r$, the resource owner (i.e., $SP_j$) constructs an access policy $P_{r,c}$, which is declared as a Boolean rule of the required attributes. $P_{r,c}$ defines a set of actions, denoted $\tau\subset\{read,write,stream\}$, that an authorized $SC$ may perform on resource $r$ in context $c$. Note that $SP_j$ has the authority to construct, update, and revoke access policies for all of its resources. $P_{r,c}$ is expressed as follows:
    \begin{equation}
    \label{eq:policy-definition}
        { P }_{ r,c }=\left< A_p , c_p, \tau_p, U_r, \varphi_r, R_{SC}^{min}, T_{SC}^{min}\right> 
    \end{equation}
    where $A_p=\{\alpha^1_p, \ldots, \alpha^n_p\}$ is a set of mandatory attributes, $c_p=\left<t, l\right>$ is a set of allowed contexts (time and access throughput limit), $U_r$ is the refresh rate of the data, $\varphi_r$ is the fee (if any) to access the resource in a cryptocurrency unit, and $R_{SC}^{min}$ and $T_{SC}^{min}$ are the minimum reputation and trust scores to access resource $r$. Note that, half of the fee $\varphi_r$ is returned back to the $SC$, if the $SC$ submits an honest feedback, as described in Section~\ref{subsec:feedback-mechanism}.
    
    A blockchain transaction ${ TX }_{ pol }$ is executed to store access policy $P_{r,c}$ to the main blockchain $MB$ and serves as the basis for the smart contract $CTR_{pol}$ that manages authorization of any $SC$. ${ TX }_{ pol }$ is expressed as follows:
    \begin{equation}
    \label{eq:tx-pol}
        { TX }_{ pol }=\left[ { P }_{ r,c }|timestamp|Sig_{ SP } \right] 
    \end{equation}
    where $timestamp$ and $Sig_{ SP }$ correspond to the timestamp of policy creation and signature of $SP$, respectively.


\subsection{Authorization Process}
\label{subsec:authorization-process}
The proposed authorization process is based on the ABAC scheme, in which access is given to users that satisfy certain attributes described in the access policy. Access control policy is enforced by $CTR_{pol}$ in $MB$ by evaluating an incoming authorization request to access resource $r$ on context $c$ based on specific Boolean attribute rule-sets in the access policy $P_{r,c}$.
Successful authorization results in issuance of an access token by $CTR_{pol}$, which can be used by the $SC$ to access the resource multiple times without repeating the authorization process.

    \subsubsection{Initial system setup}
    We presume that there is a trusted smart city regulator that initially deploys the two core smart contracts (i.e., $CTR_{TRS}$ and $CTR_{pol}$) to the main blockchain $MB$. In addition, independent $AA$s create and maintain their own $pb$, acting as sidechains, and announce their presence. Each $AA_k$ deploys its own Attribute Provider Contract ($CTR^k_{AP}$) to $pb_k$ and actively listens for incoming attribute lookup events from $CTR_{pol}$ on $MB$.

    \begin{figure}
    \centering
    \includegraphics[width=0.47\textwidth]{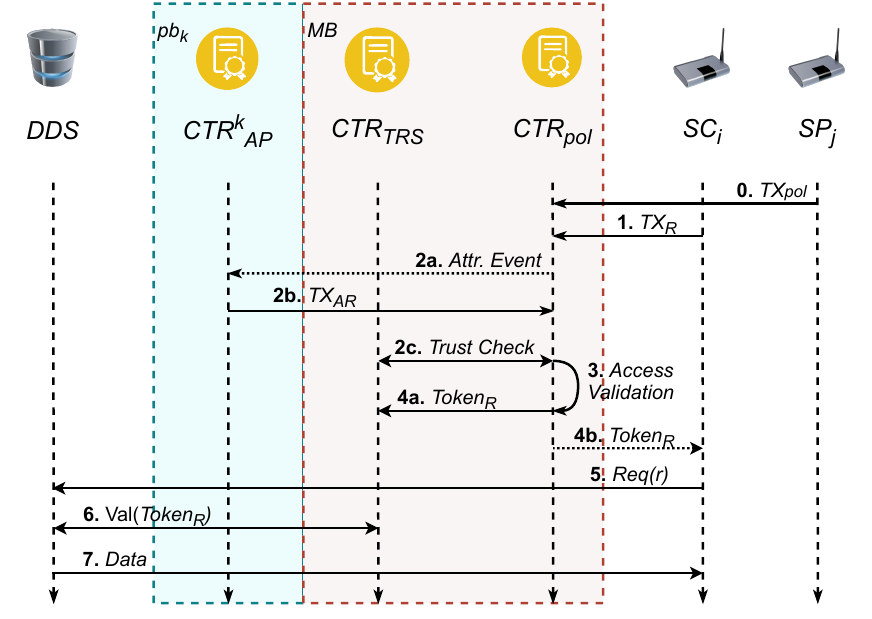}.
    \caption{Authorization process.}
    \label{fig:authorization-process}
    \end{figure}

    \subsubsection{Steps in authorization}
    The authorization process is shown in Fig.~\ref{fig:authorization-process}. Prior to authorization, $SC_i$ must have its attributes registered to one of any available $AA$s. We assume that resource related information is already known in advance and stored in $DDS$.
    
    \begin{algorithm}
    \caption{Access Request Validation}
    \label{alg:access-validation}
    \begin{algorithmic}[1]
    \renewcommand{\algorithmicrequire}{\textbf{Input:}}
    \renewcommand{\algorithmicensure}{\textbf{Output:}}
    \REQUIRE $P_{r,c}$, $TX_R$, $T^{SP_j}_{SC_i}$, $R_{SC_i}$, and $A_i$
    \ENSURE  $Token_R$ \OR $\emptyset$
    \STATE $authorized \leftarrow 0 $
        \IF{$A_p \subset A_i$ \AND $\tau \subset \tau_p$}
            \IF{$T^{SP_j}_{SC_i} \ge T^{min}_{SC}$ \AND $R_{SC_i} \ge R^{min}_{SC}$}
                \IF{$getBalance(PK_{SC_i}) \ge \varphi_r$}
                    \STATE $authorized \leftarrow True$
                \ENDIF
            \ENDIF
        \ENDIF
    \IF{$authorized$}
        \RETURN $Token_R \leftarrow \left<Exp_R,l,t\right>$
    \ELSE
        \RETURN $\emptyset$
    \ENDIF
    \end{algorithmic}
    \end{algorithm}

    \begin{itemize}
        \item \textit{Initialization (Step 0)}: First, $SP_j$ defines access policy $P_{r,c}$ and then initiates a transaction $TX_{pol}$ to store the access control policy to $MB$.
        
        \item \textit{Step 1}: $SC_i$ authorizes itself to $CTR_{pol}$ by initiating transaction $TX_R$, which is defined as:
        \begin{equation}
        \label{eq:tx-r}
            TX_R = \left[ r | \tau | Sig_{SC_i} \right]
        \end{equation}
        where $r$ is the resource identifier, $\tau$ is the requested action, and $Sig_{SC_i}$ is the signature of $SC_i$ used for authentication.
        
        \item \textit{Step 2}: Step 2a and 2b correspond to the bridging mechanism,  which $CTR_{pol}$ uses to validate $SC_i$'s attributes. In Step 2a, $CTR_{pol}$ emits an attribute validation event to the blockchain. Each $pb$ connected to $MB$ listens to the events and checks if $PK_{SC_i}$ is registered on their chain. One of the $CTR^k_{AP}$'s will find $PK_{SC_i}$ on $pb_k$. This $CTR^k_{AP}$ validates the attributes and returns the result as an attribute response transaction $TX_{AR}$ to $CTR_{pol}$ (Step 2b). In addition, $CTR_{pol}$ also calls $CTR_{TRS}$ to obtain $T^{SP_j}_{SC_i}$ and $R_{SC_i}$ (Step 2c).
        
        \item \textit{Step 3}: $CTR_{pol}$ executes Algorithm~\ref{alg:access-validation} to validate whether $SC_i$ satisfies the required attributes and reputation threshold on $P_{r,c}$ and has sufficient balance to pay the access fee.
        
        \item \textit{Step 4}: Upon successful access validation, $CTR_{pol}$ issues $Token_R$, which is defined as:
        \begin{equation}
            Token_R=\left<{Exp}_{R}, l, t\right>
        \end{equation}
        where $Exp_R$ is the token expiration time, $l$ is the rate limit, and $t$ is token timestamp. $CTR_{pol}$ sends $Token_R$ to $CTR_{TRS}$ to update the trust and reputation score of $SC_i$ (Step 4a) and to $SC_i$ as a proof that $SC_i$ has been authorized to access resource $r$ (Step 4b). In addition, $CTR_{pol}$ sends half of the fee $\varphi_r$ to $SP_j$'s account and the other half of $\varphi_r$ to $CTR_{TRS}$ for feedback reward, as described in Section~\ref{subsec:feedback-mechanism}.
        
        \item Step 5: Based on the specification of resource $r$, $SC_i$ locates the corresponding $DDS$ to access the data.
        $SC_i$ initiates the process by submitting a request to obtain a nonce from $DDS$. Subsequently, $SC_i$ sends $Req(r)$ message to $DDS$ over a secure channel that contains $Token_R$, the cryptographic nonce, and the signature of $SC_i$.
        
        \item Step 6: To prevent access token forgery, $DDS$ validates $Token_R$ to $CTR_{TRS}$. $DDS$ also checks if an $SC$ sends quick successive requests at a rate higher than a certain rate limit $l$ to prevent DoS attack. The use of forged token and a DoS attack would result in violation of access policy and would be reported to $CTR_{TRS}$, hence a drop in $T_{SC_i}^{SP_j}$. In addition, we assume that a dedicated intrusion detection mechanism exists to mitigate DoS attacks, e.g., blacklisting the node.
        
        \item Step 7: Upon successful validation, $DDS$ responds to the access request by sending the data signed by the $SP$ and an access timestamp signed by the $DDS$.
    \end{itemize}
    
    \begin{algorithm}
    \caption{Feedback Mechanism}
    \label{alg:feedback-mechanism}
    \begin{algorithmic}[1]
    \renewcommand{\algorithmicrequire}{\textbf{Input:}}
    \renewcommand{\algorithmicensure}{\textbf{Output:}}
    \REQUIRE $P_{r,c}$, $TX_F$
    \ENSURE  $True$ \OR $False$
    \STATE $last\_update \leftarrow getUpdateTimestamp(Data)$
    \STATE $access\_timestamp \leftarrow getAccessTimestamp(Data)$
    \STATE $evidence \leftarrow checkSig(Data)$
    \STATE $result \leftarrow False$
    \IF{$H(Token_R)$ exist}
        \RETURN $result$
    \ENDIF
    \IF{$(access\_timestamp - last\_update) < U_r$}
        \IF{$F^{SC_i}_{SP_j, r} = positive$ \AND $evidence = True$}
            \STATE $\varepsilon_t \leftarrow \varepsilon_{pos}$
            \STATE $result \leftarrow True$
        \ENDIF
    \ELSE
        \IF{$F^{SC_i}_{SP_j, r} = negative$ \AND $evidence = True$}
            \STATE $\varepsilon_t \leftarrow \varepsilon_{neg}$
            \STATE $result \leftarrow True$
        \ENDIF
    \ENDIF
    \IF{$result = True$}
        \STATE $reCalculate\ T^{SC_i}_{SP_j},A_{SP_j}, R_{SP_j}$
        \STATE $sendCryptoTo(SC_i)$
    \ELSE
        \STATE $\delta_t \leftarrow \delta_{neg}$
        \STATE $reCalculate\ T^{SP_j}_{SC_i},A_{SC_i},R_{SC_i}$
        \STATE $sendCryptoTo(SP_j)$
    \ENDIF
    \RETURN $result$
    \end{algorithmic}
    \end{algorithm}
    
    \subsubsection{Asynchronous Authorization}
    Unlike typical authorization mechanisms which require both $SP$ and $SC$ to be active and connected to the system simultaneously~\cite{Ding2019}, the nature of our proposed authorization mechanism allows $SP$ to be offline when $SC$ is requesting for an access. In Step 0, $SP_j$ initiates $TX_{pol}$ to indicate that a resource is available to access with an access policy of $P_{r,c}$. Step 0 is actually an implicit delegation process, in which $SP_j$ delegates the authorization process to $CTR_{pol}$. As such, while an $SC_i$ is requesting an access, $SP_j$ may be offline temporarily, for instance, to save energy which is typical in IoT. We argue that this kind of asynchronous authorization, in which all participating nodes are not required to be synchronously active at the same time, would offer greater flexibility for the $SP$.

\begin{figure*}
    \centering
    \begin{subfigure}[b]{0.32\textwidth}
        \includegraphics[width=\textwidth]{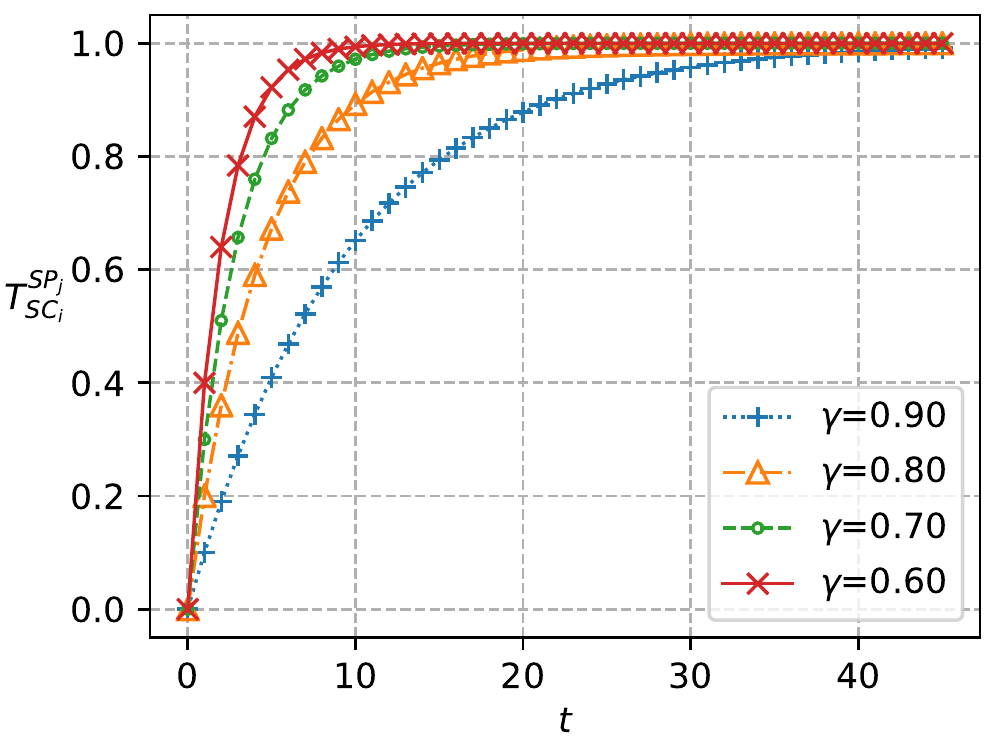}
        \caption{Convergence of trust score $T^{SP_j}_{SC_i}$ with different $\gamma$ ($\delta_{pos} = 1$ and $\delta_{neg} = -3$)}
        \label{fig:trust-score-graph}
    \end{subfigure}
    ~
    \begin{subfigure}[b]{0.32\textwidth}
        \includegraphics[width=\textwidth]{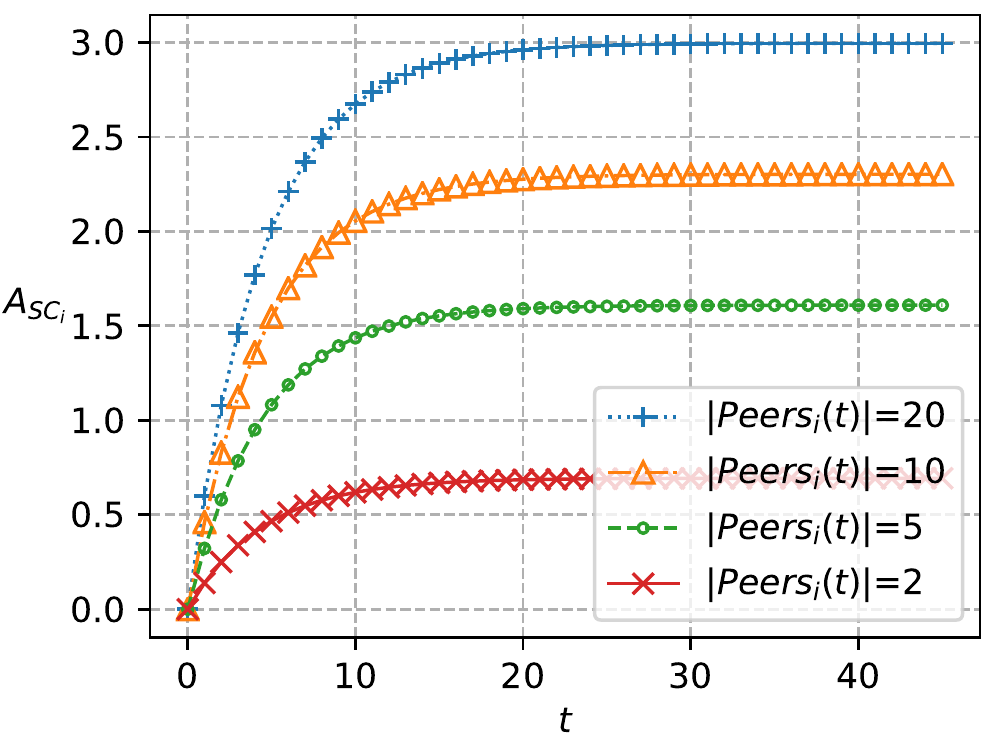}
        \caption{Aggregation of trust score $A_{SC_i}$ with different $|Peers_i(t)|$}
        \label{fig:aggregation-score-graph}
    \end{subfigure}
    ~
    \begin{subfigure}[b]{0.32\textwidth}
        \includegraphics[width=\textwidth]{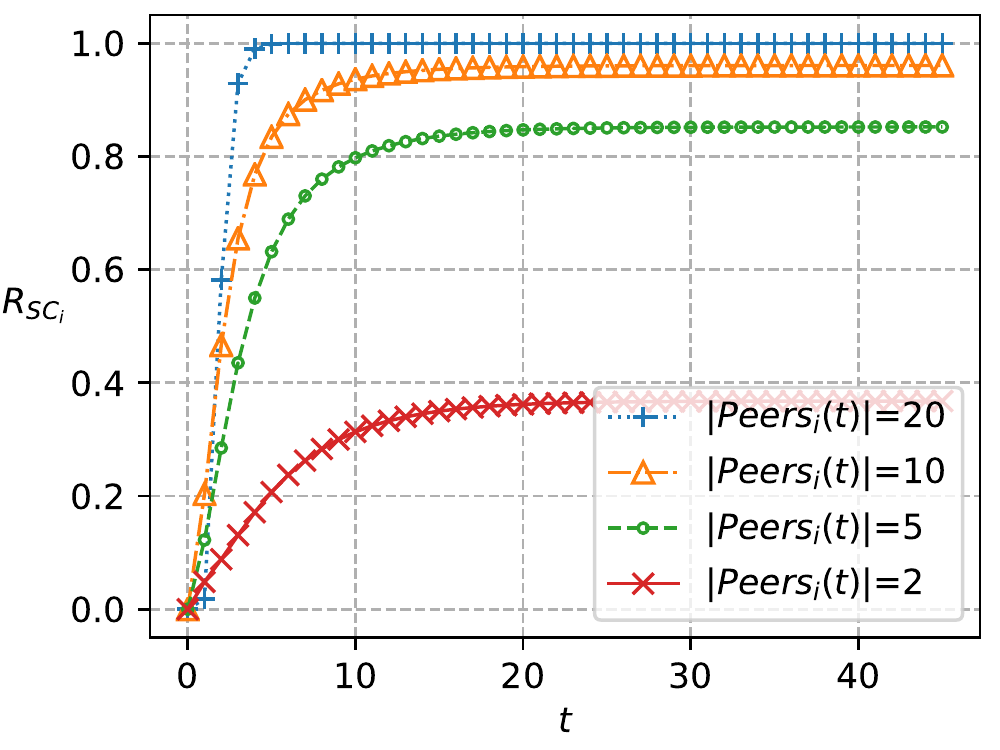}
        \caption{Reputation score $R_{SC_i}$ with different $|Peers_i(t)|$ ($a = 1$, $b = -4$, and $c = -2$)}
        \label{fig:reputation-score-graph}
    \end{subfigure}
    \caption{The convergence of trust and reputation score over time with varying parameters for all positive interactions.}
    \label{fig:trust-and-reputation-evolution}
\end{figure*}

\begin{figure*}
    \centering
    \begin{tabularx}{\linewidth}{XXX}
        \includegraphics[width=0.32\textwidth]{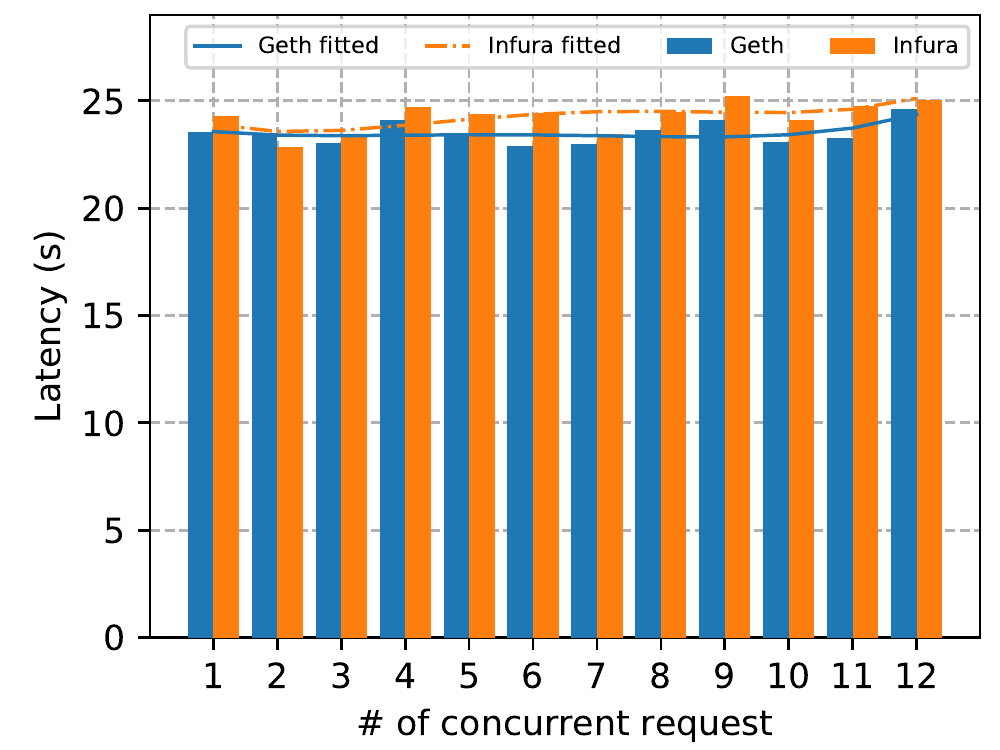}
        \caption{Comparison of authorization latency via Geth light client and Infura API Gateway (Step 1-7).}
        \label{fig:authorization-latency-bar-chart}
        &
        \includegraphics[width=0.32\textwidth]{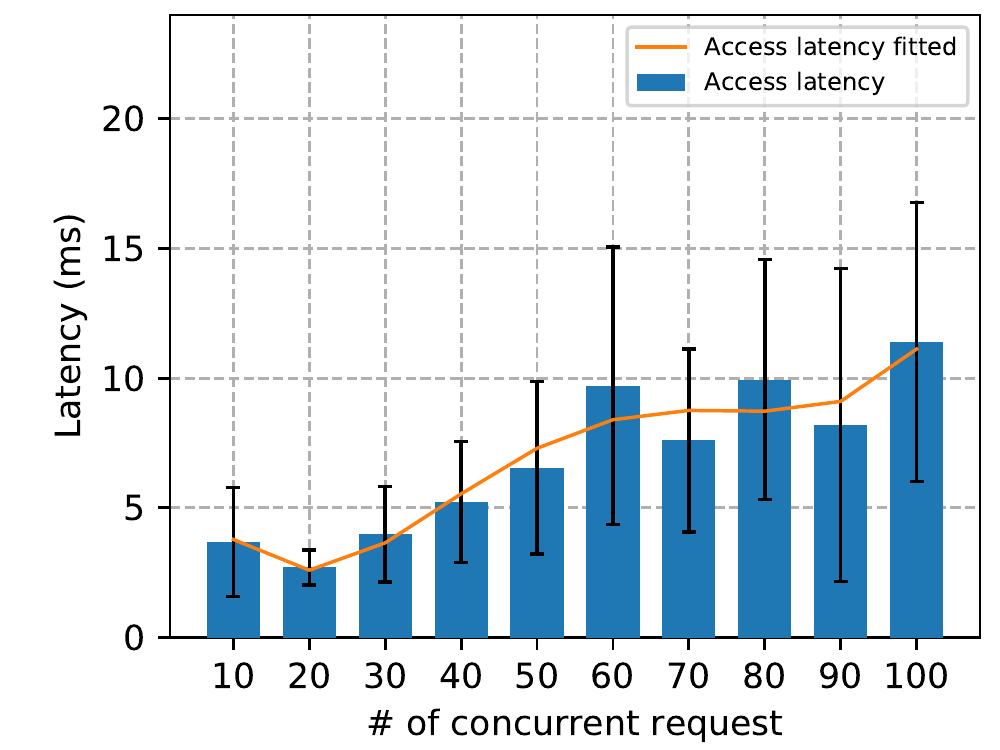}
        \caption{Comparison of access latency, in which $SC$ has already obtained $Token_R$ beforehand (Step 5-7).}
        \label{fig:access-latency-bar-chart}
        &
        \includegraphics[width=0.32\textwidth]{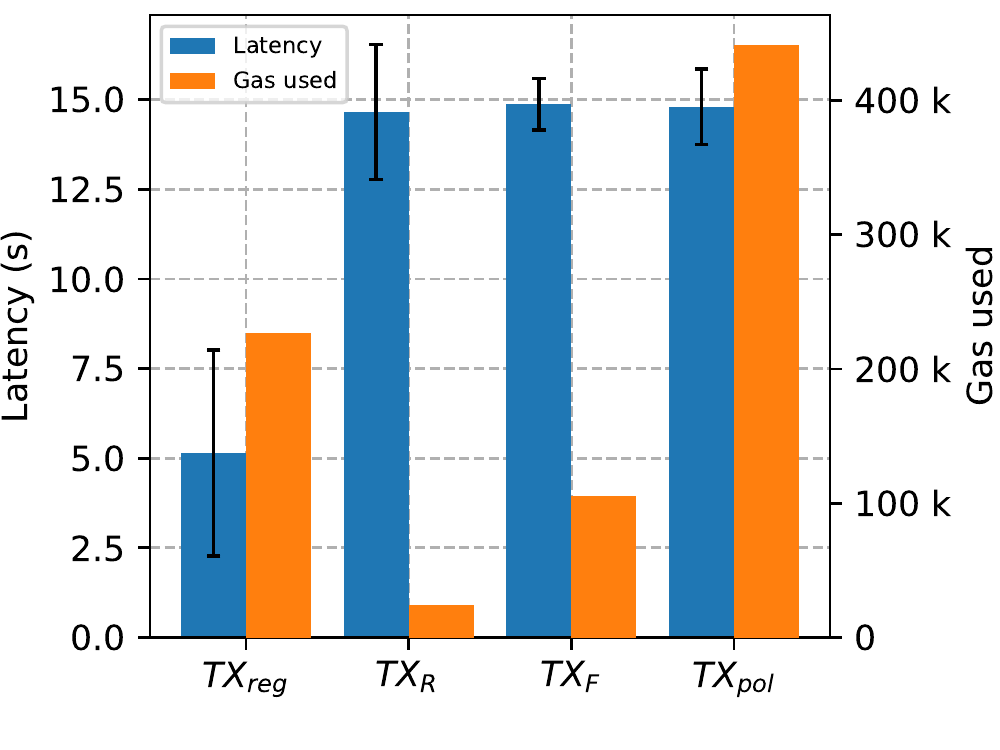}
        \caption{Latency and required gas of attribute registration, authorization, feedback, and policy registration.}
        \label{fig:gas-used}
    \end{tabularx}
    \vspace{-2\baselineskip}
\end{figure*}

\subsection{Feedback Mechanism}
\label{subsec:feedback-mechanism}
    While $T^{SP_j}_{SC_i}$ and $R_{SC_i}$ are updated during the authorization process and data access, the trust and reputation score of $SP_j$ (i.e., $T^{SC_i}_{SP_j}$ and $R_{SP_j}$) are updated when $SC_i$ submits a feedback $TX_F$ to $CTR_{TRS}$ after accessing resource $r$, which is defined as:
    \begin{equation}
        TX_F = \left[ F^{SC_i}_{SP_j, r} | Data | H(Token_R) | Sig_{SC_i} \right]
    \end{equation}
    where $F^{SC_i}_{SP_j, r}$ is the binary feedback from $SC_i$ after accessing resource $r$(i.e., either positive or negative), $Data$ is the obtained data with last update and access timestamps signed by $SP_j$ and $DDS$,
    $H(Token_R)$ is the hash of $Token_R$, and $Sig_{SC_i}$ is the signature of $SC_i$ used for authentication.
    Although $SC$ may access $r$ multiple times, $TX_F$ can only be submitted once for each $Token_R$ and $CTR_{TRS}$ will always check for duplicated feedback. To motivate $SC_i$ to provide feedback, $SC_i$ receives half of $\varphi_r$ as a reward for submitting an honest feedback.
    
    Recall that as defined in $P_{r,c}$, $SP_j$ is expected to periodically update the data according to $U_r$. Positive $F^{SC_i}_{SP_j, r}$ refers to timely data, while the negative refers to obsolete data that has not been updated accordingly. $CTR_{TRS}$ validates $TX_F$ by comparing the last update and access timestamps against $U_r$ and verifying the signature of $SP_j$ and $DDS$ as supporting evidence. The signature based evidence prevents the $SC_i$ to forge the timestamps for malicious purposes. $CTR_{TRS}$ inspects the evidence and increases the trust and reputation score of $SP_j$ when the evidence supports the feedback. On the contrary, $CTR_{TRS}$ will drop $SC_i$'s trust and reputation score if $SC_i$ submits misleading feedback. We show the procedures of our feedback mechanism in Algorithm~\ref{alg:feedback-mechanism}.


\begin{figure*}
    \centering
    \begin{tabularx}{\linewidth}{XXX}
        \includegraphics[width=0.32\textwidth]{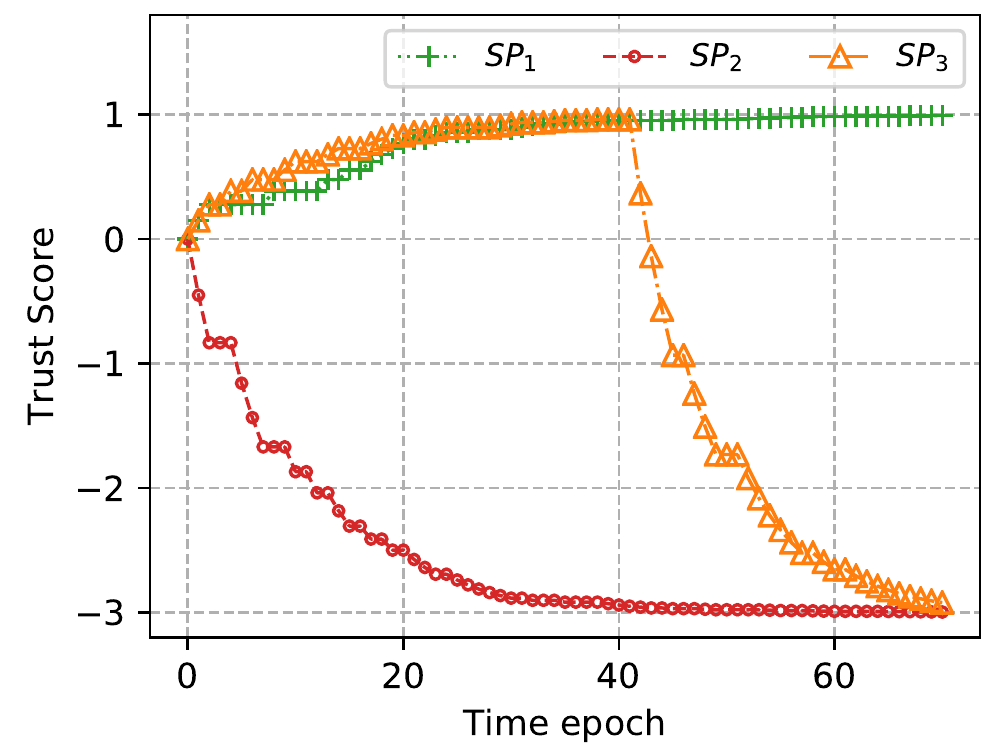}
        \caption{The trust evolution comparison of honest and malicious $SP$ ($\mu = 0.8$, $\varepsilon_{pos} = 1$, and $\varepsilon_{neg} = -3$).}
        \label{fig:trust-honest-malicious}
        &
        \includegraphics[width=0.32\textwidth]{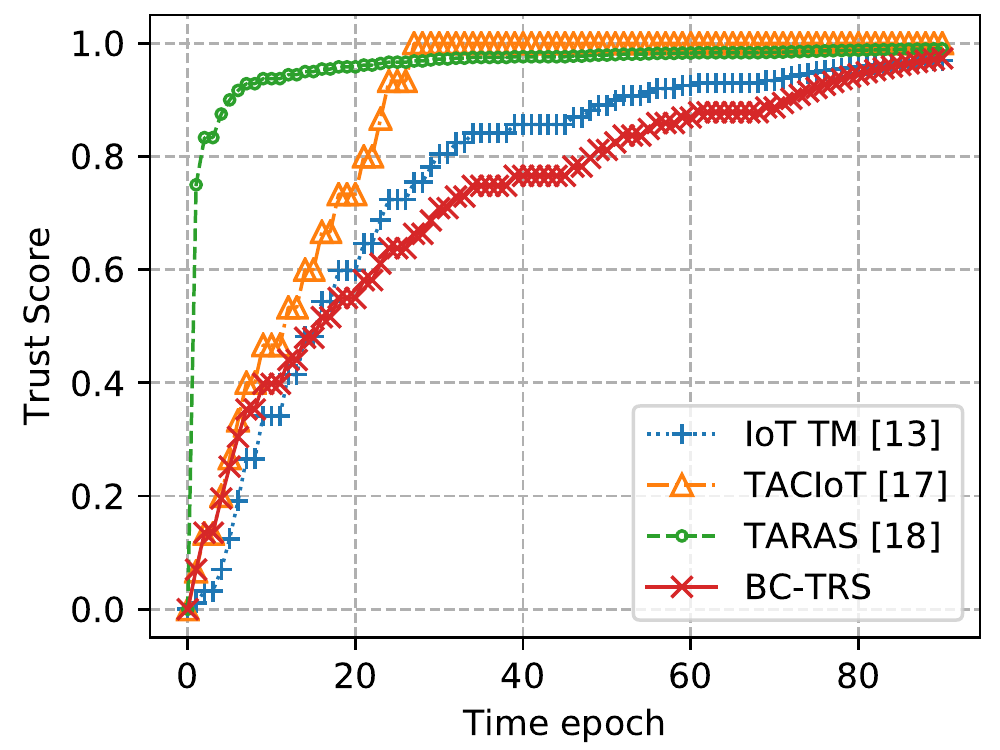}
        \caption{The comparison of trust score convergence between several proposals  for positive interactions.}
        \label{fig:compare-positive}
        &
        \includegraphics[width=0.32\textwidth]{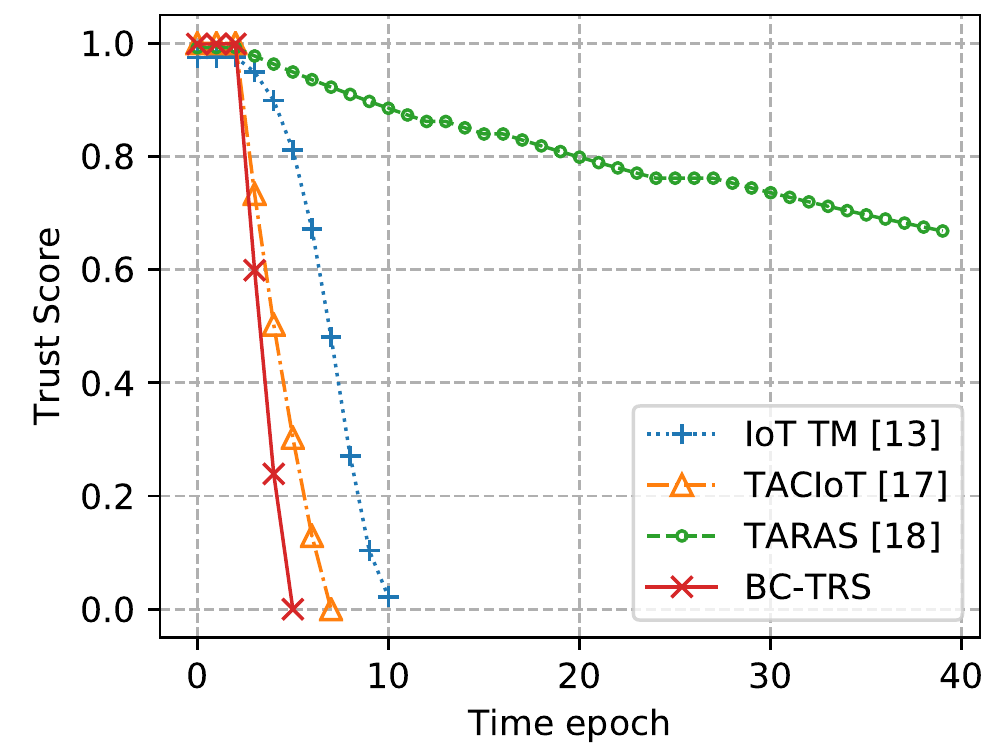}
        \caption{The comparison of trust score convergence between several proposals for negative interactions.}
        \label{fig:compare-negative}
    \end{tabularx}
    \vspace{-2\baselineskip}
\end{figure*}

\section{Performance Evaluation}
\label{sec:performance-evaluation}
In this section, we present the performance evaluation of our solution based on our proof-of-concept implementation, that was tested on a public Ethereum test network. We evaluate our solution on reputation specific metrics, namely trust and reputation evolution and blockchain performance metrics such as, authorization latency and gas used.

\subsection{Proof-of-Concept Details}
\label{subsec:poc}
    We selected Ethereum as the blockchain platform for the proof-of-concept implementation, primarily due to three reasons. First, Ethereum is suitable for both permissionless and permissioned environment, which is necessary for our proposed system.
    In the Ethereum permissionless environment, each peer may freely join and leave the network at any given time and there is no authority that administers user registration. Note that, each peer must use the elliptical curve SECP-256k1~\cite{wood2014ethereum} key generation approach used in Ethereum to obtain valid credentials.
    Second, Ethereum supports smart contracts, written in Solidity, which is a Turing complete programming language~\cite{solidity066}.
    The smart contract is executed within a decentralized Ethereum Virtual Machine (EVM), which acts as a trusted platform for decentralized computation~\cite{wood2014ethereum}.
    Third, Ethereum offers an in-built cryptocurrency for token transactions, with Ether being the default fundamental token for transactions. We tested our smart contracts on the Rinkeby public Ethereum test-network, which implements Proof-of-Authority as the consensus mechanism.
    Note that, our proposed solution can also be implemented on any blockchain platform that supports smart contract execution, for instance, Hyperledger Fabric~\cite{10.1145/3190508.3190538}.
    
    To implement our proposed framework, we built a lab-scale testbed which consists of six Raspberry Pi 3 B+ (1GB RAM, 1.4GHz 64-bit quad-core ARM CPU, Raspbian 10 buster) as IoT nodes and a single MacBook Pro 2019 (8GB RAM, 1.4 GHz quad-core Intel Core i5 CPU, macOS 10.14.6) acting as both $AA$ and $DDS$. MacBook Pro computer and all Raspberry Pis ran \texttt{geth} v1.9.12 as an Ethereum client to connect to the Rinkeby test-network.
    Note that, due to the limited resources and computation capabilities, Raspberry Pis only ran a geth light client that synchronizes the blockchain by only downloading and verifying the block headers, without executing any transactions or retrieving any associated state.
    We used Python v3.7.7 and bash scripts to simulate interactions among nodes with \texttt{web3py} v5.11 and \texttt{py-solc} v3.2.0 library for communicating with Ethereum peers and compiling the smart contracts, respectively. We wrote our implementation of $CTR_{pol}$, $CTR_{TRS}$, and $CTR_{AA}$ in Solidity v0.6.6~\cite{solidity066}, with native support for efficient computation and verification of hashes~\cite{8752021}.
    

\subsection{Evaluation Results}
\label{subsec:results}
    
    \subsubsection{Trust and Reputation Convergence}
    We study the convergence of our proposed TRS by varying the weighting parameters and present the results in Fig.~\ref{fig:trust-and-reputation-evolution}. By varying the ageing parameters (i.e., $\gamma$ and $\mu$), we basically allocate different weights for recent and older interactions in calculating the trust score. Fig.~\ref{fig:trust-score-graph} shows the convergence of $T^{SP_j}_{SC_i}$ with different $\gamma$ for all positive interactions with $\delta_{pos} = 1$ and $\delta_{neg} = -3$. We can observe that higher values of $\gamma$ deliver more gradual growth in trust score evolution. However, all lines converge to $1$, which is the maximum value. Recall that, in Eq.~\eqref{eq:trust-boundaries}, the value of $max(T^{SP_j}_{SC_i})$ is capped at $\delta_{pos}$.
    Recall in Section~\ref{sec:trust-and-reputation} that the reputation score of a node (i.e., $R_{SC}$ or $R_{SP}$) is affected by how many unique peers the node has interacted with. Intuitively, for positive interactions, having more peers would increase the reputation score. As shown in Fig~\ref{fig:aggregation-score-graph}, $A_{SC_i}$ is directly proportional to $\ln |Peer_i(t)|$.
    In Fig~\ref{fig:reputation-score-graph}, we can see that although the reputation score $R_{SC_i}$ is positively correlated with $|Peer_i(t)|$, the reputation score $R_{SC_i}$ is capped at 1, which corresponds to the variable $a$ in Eq.~\eqref{eq:reputation-sc-gompertz} and~\eqref{eq:reputation-sp-gompertz}.
    
    \subsubsection{Authorization and access latency}
    We compare the authorization latency, i.e., the time taken from Step 1 to Step 7 in Fig~\ref{fig:authorization-process}, for two different connection methods. First, an $SC$ connects to $MB$ to authorize itself by running a light geth node. Second, an $SC$ connects to the blockchain via Infura\footnote{\url{https://infura.io/}}, a third-party API provider. Note that the fundamental difference in both connection schemes is about main blockchain synchronization, which is required in geth but not in Infura.
    Since, the IoT nodes sign the transactions locally to ensure the security of their private keys, it implies that Infura provider does not have access to the private keys of the nodes.
    We repeated the experiment 30 times with different number of concurrent requests and plot the results in Fig.~\ref{fig:authorization-latency-bar-chart}. In general, authorization via geth achieves slightly lower latencies than via Infura. We can observe that increasing number of concurrent requests results in no significant increase of the authorization latency.
    We also examine the access latency (i.e., Step 5-7), in which the $SC$ re-uses the $Token_R$ to access $r$ before the token expires. We measured the latency for different number of concurrent requests and repeated the experiments 30 times. As shown in Fig.~\ref{fig:access-latency-bar-chart}, the access latency is three orders of magnitude lower than the authorization latencies, as the $SC$ is not required to repeat Step 1-4. However, there is an increasing trend when the number of concurrent requests is increased.

    \subsubsection{Latency and required gas}
    In Ethereum, there is a fee to execute transactions that alter the blockchain state. The fee, which is referred to as Gas, depends on the number of EVM opcodes involved during the execution of a particular function~\cite{wood2014ethereum}. The gas also helps to avoid excessive execution of smart contracts, e.g., infinite loops. In practice, the gas is relatively small and sometimes negligible.
    
    We investigate the amount of gas required to execute essential functions in our smart contract design, namely attribute registration ($TX_{reg}$), authorization ($TX_R$), feedback ($TX_F$), and policy registration ($TX_{pol}$). In addition, we investigate the latency to execute these functions. We repeated the experiments 30 times and plot the results in Fig.~\ref{fig:gas-used}. The transaction latencies are similar for $TX_R$, $TX_F$, and $TX_{pol}$, as these transactions are executed in $MB$. The transaction latency for $TX_{reg}$ is relatively lower as it is executed in $pb$, which has faster block generation time. $TX_R$ requires the least Gas, while $TX_{pol}$ requires the most, due to the different amount of data contained within the $TX_R$ and $TX_{pol}$, which in turn results in a different number of EVM opcodes in the execution (refer to Eq.~\eqref{eq:policy-definition}~\eqref{eq:tx-pol}~\eqref{eq:tx-r}).
    
    \subsubsection{Trust evolution of honest and malicious node}
    To study how the convergence of trust scores, we simulate three $SP$s with different behavior. One of the $SP$s is honest for all $t$, another is malicious for all $t$, while the third is an $SP$ that initially acts honestly until $t=40$, then becomes malicious for the rest of the simulation. We use $\mu = 0.8$, $\varepsilon_{pos} = 1$, and $\varepsilon_{neg} = -3$ as the parameters and plot the results in Fig~\ref{fig:trust-honest-malicious}. We can see that all $SP$ scores start at $0$ as the initial trust value. The trust scores for honest and malicious $SP$s reach the maximum and minimum boundaries approximately at similar time $t=40$. Note that the maximum and minimum trust scores follow equation~\eqref{eq:trust-boundaries}. At $t>40$, the trust score of the third $SP$ drops significantly. As the interactions continue consistently, the trust scores remain unchanged.

\begin{table}
\centering
\caption{Comparing the complexity of trust managements for IoT authorization.}
\label{ta:trust-mgmt-comparison}
\begin{tabular}{lcc}
\toprule
\multirow{2}{*}{\textbf{Proposal}} & \multicolumn{2}{c}{\textbf{Complexity}} \\
 & \textit{Trust Computation} & \textit{Storage} \\
\midrule
TACIoT~\cite{BernalBernabe2016} & $O(p * j)$ & $O(1)$ \\
TARAS~\cite{Gwak2018} & $O(s_i)$ & $O(1)$ \\
IoT TM~\cite{icbc2020} & $O(p)$ & $O(\sum_{x=1}^{N} p_x)$ \\
BC-TRS & $O(1)$ & $O(1)$ \\
\bottomrule
\end{tabular}
\end{table}
    
\subsection{Comparative Analysis}
\label{subsec:numerical-comparison}
As a quantitative comparison, we examine the trust evolution of our proposed solution against other proposals in trust management for IoT authorization~\cite{icbc2020,BernalBernabe2016,Gwak2018}. We simulated two nodes (i.e., $SC$) which have different behavior and calculated the trust score of each node using different trust models. Note that in each trust model, we set the parameters as recommended in the corresponding paper, i.e., $a=1$, $b=6$, $c=0.1$, $\gamma=0.95$, $\delta_{pos}=3$, and $\delta_{neg}=-5$ for~\cite{icbc2020}; $S=7$, $c^p_j=0.5$, and $N=15$ for~\cite{BernalBernabe2016}; and $W=1$ for~\cite{Gwak2018}. We plot the trust score evolution of benign and malicious interactions in Fig.~\ref{fig:compare-positive} and Fig.~\ref{fig:compare-negative}, respectively. For honest $SC$s, we see that although all trust models converge to similar upper boundary, each trust model has different convergence rate. TARAS seems to have the quickest convergence rate at which it is able to reach 0.8 at less than 10 time epoch, while BC-TRS reaches 0.8 at time epoch around 50. Note that a faster convergence rate is generally less preferred as it is more vulnerable to newcomer attacks. On the other hand, a fast decline is preferred to penalize malicious nodes. Fig.~\ref{fig:compare-negative} shows that three models demonstrate a significant abrupt decline for malicious $SC$s, in which the trust score drops to zero within 10 time epoch. We note that BC-TRS reaches zero only within 3 malicious interactions.

In Table~\ref{ta:trust-mgmt-comparison}, we compare our proposed solution against other related trust management models for IoT authorization. We see that in terms of trust computation, the complexity of other models depends on certain parameters, such as the number of previous interaction ($p$), trust dimension ($j$), and group member ($s_i$). Recall in Section~\ref{sec:trust-and-reputation} that our trust score can be calculated using simple recursion, which reduces our trust computation complexity to $O(1)$. In TACIoT and TARAS, the IoT devices rely on a centralized trust manager to store interaction history, which reduces the storage requirement to $O(1)$. IoT TM, however, requires each $SP$ to store the interaction history with any $SC$ in its own storage, which amounts to the number of service consumer ($N$) and the number of previous interaction $p_x$ for each $SC$. Our proposed solution offloads the storage to blockchain, which reduces the storage complexity to $O(1)$. Our proposed solution also offers privacy preservation by storing sensitive information on private sidechains. Our solution supports asynchronous authorization, in which both $SP$ and $SC$ are not required to be simultaneously online to perform authorization. Lastly, our model also supports bidirectional trust assessment, wherein each $SP$ can assess the trustworthiness of $SC$ and vice versa, as seen in Fig.~\ref{fig:trust-model}.

\section{Discussion}
\label{sec:discussion}

\subsection{Reliability and Scalability}
\label{subsec:reliability-scalability}

We design our proposed solution to be resilient against attacks discussed in Section~\ref{subsec:threat-model}.
When an adversary launches bad mouthing attacks, the request signatures confirm the authenticity of the sender, which prohibits the adversary from stealing or forging invalid tokens of other nodes for maliciously reducing its reputation score. Our framework is also resilient to Sybil and newcomer attacks, which are handled by the attribute registration mechanism. Any participant cannot re-register itself to $AA$, as the $AA$ keeps track of the attributes relation to the technical specifications and ownership information of the device. In addition, our trust computation model prevents adversaries to illegitimately increase their reputation, i.e., performing ballot-stuffing and self-promoting attacks, as our model also considers the number of peers a node has interacted with to calculate the reputation score (see Section~\ref{sec:trust-and-reputation}).

In our solution, blockchains serve as the backbone of the network. Subsequently, the scalability of our system is inherited from the underlying blockchain instantiation. Note that our paper does not address the issue of blockchain scalability, but there have been active contributions from the community about possible solutions to limited scalability~\cite{8962150}. Fig.~\ref{fig:authorization-latency-bar-chart} indicates that our solution can achieve a stable latency for different number of concurrent requests. However, a slightly increasing trend is observed in Fig.~\ref{fig:access-latency-bar-chart}, which indicates that $DDS$ might introduce a bottleneck in the network and hinder scalability. Note that this trend is observed as we implemented $DDS$ on a single node, in fact a possible solution to remove the bottleneck is to add redundancy in $DDS$ via a scalable data storage~\cite{Ali:2017:IDP:3131542.3131563}. 

As shown in Fig. 3, our authorization process involves three main stages, e.g., attribute validation, trust computation, and access validation, which has different computational complexity in each stage. First, in attribute validation, $CTR^k_{AP}$ may perform binary search to find the attributes of an $SC$, which results in a complexity of $O(\log n)$, where $n$ is the number of registered $SC$s. Second, our trust model involves a simple recursion in trust computation, which only requires $CTR^k_{TRS}$ to either increment or decrement the previous trust value (complexity of $O(1)$). Third, access validation involves iteratively comparing the required attributes in $P_{r,c}$ against $SC$'s attributes, which has a computational complexity of $O(n)$. In general, the overall complexity of the authorization process is linear ($O(n)$).

\subsection{Implications of TRS for Theory and Practice}
\label{subsec:implications-trs}
It is known that access control demands highly sensitive consideration that requires certainty of who can access what resource. However, trust-based approaches are generally probabilistic in nature, which may make them vulnerable to exploitation. Relying entirely on a trust score for access control is risky, as attackers could conceivably build up trust to gain access. To reconcile these issues, we based our trust-based approach on an established attribute based access control scheme with trust and reputation score as auxiliary attributes. We argue that incorporating explicit trust scores would help to achieve a dynamic and flexible access control system while also prohibiting access to malicious and compromised nodes.

The evaluation results indicate that the authorization process incurs appreciable latency. However, the latency of re-accessing a resource with a previously obtained token is relatively low. Note that, in general $SC$s may only need to authorize themselves once $Token_R$ has expired (defined in $Exp_R$). One possible application of our proposed model is in managing access to critical infrastructure, such as power grid or water supply network, that require stringent measures of authorization, limited to highly trusted actors with explicit inherent attributes. For instance, we can implement our solution to quantitatively oversee the performance of smart building contractors. Ideally, smart building contractors that regularly maintain critical infrastructure should be highly trusted, demonstrated by an exceptional track record of prior interactions. The contractors, after deploying IoT devices on the infrastructure premises, would either act as a $SP$ or $SC$ depending on the specifications. The contractors regularly monitor the infrastructure condition by frequently checking IoT sensor readings and reporting certain anomalies to corresponding authorities. Violations in the maintenance procedures, i.e., malicious actions, would result in rigorous penalty as per our proposed TRS.

\subsection{Limitations and Future Work}
\label{subsec:limitations-future-work}
Our proposed solution has a few limitations, which we aim to solve in the future work. First, we may encounter a bootstrapping problem, in which a new $SC$ with zero reputation score cannot participate in the network. However, we presume that some $SP$s would allow nodes with zero score to access their resource, through which a newly joined node may initially build up its reputation score.
We also aim to circumvent this limitation by designing an endorsement-based access policy, in which new $SC$s with zero reputation score may request an access of a resource with a prior endorsement with another $SC$ with adequate reputation score.
Second, while our TRS captures attacks and violations against access control, violations in the attribute registration process in the sidechains cannot be mitigated. As an implication, a malicious $SC$, which initially tries to register itself using invalid attributes, would have a similar initial score of $T^{SP_j}_{SC_i}$ with another honest $SC$ when they join the network for the first time.
We plan to address this issue by developing a more robust multidimensional trust model, in which multiple dimensions of metrics are taken into account, including violations in the attribute registration process and intrusion related information from the deployed intrusion detection system.
Another interesting aspect for future work is to implement obfuscation in the trust model to avoid de-anonymization in the blockchain.

\section{Conclusion}
\label{sec:conclusion}
In this paper, we proposed a trust-based access control framework for decentralized IoT network. We design an auxiliary TRS as part of a blockchain-based ABAC mechanism that incorporates trust and reputation scores as additional attributes for achieving dynamic and trustworthy access control mechanism. We design our framework to be blockchain agnostic, which can be implemented in any blockchain platforms that have adequate support for smart contract execution. We implemented a proof-of-concept in a public Rinkeby Ethereum test-network interconnected with a lab-scale testbed. Experimental results show that our proposed framework achieves consistent processing latencies and is feasible for implementing effective access control in decentralized IoT networks.

\section*{Acknowledgments}
The authors acknowledge the support of the Commonwealth of Australia and Cyber Security Cooperative Research Centre for this work.

\ifCLASSOPTIONcaptionsoff
  \newpage
\fi

\bibliographystyle{IEEEtran}
\bibliography{./bibtex/IEEEabrv,./bibtex/ref}

\vspace{-3\baselineskip}

\begin{IEEEbiography}
    [{\includegraphics[width=1in,height=1.25in,clip,keepaspectratio]{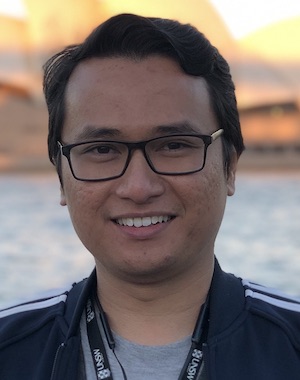}}]{Guntur Dharma Putra}
    received his bachelor degree in Electrical Engineering from Universitas Gadjah Mada, Indonesia, 2014. He received his master's degree in Computing Science from the University of Groningen, the Netherlands, 2017. He is currently a Ph.D. candidate in the School of Computer Science and Engineering, University of New South Wales (UNSW), Sydney, Australia. His research interest covers blockchain applications for securing IoT. Guntur is a student member of the IEEE.
\end{IEEEbiography}

\vspace{-3\baselineskip}

\begin{IEEEbiography}
    [{\includegraphics[width=1in,height=1.25in,clip,keepaspectratio]{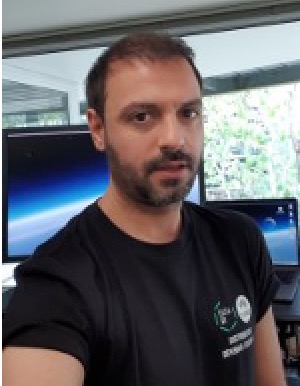}}]{Volkan Dedeoglu}
    is currently a postdoctoral research fellow in the Distributed Sensing Systems Group of CSIRO Data61.
    Volkan also holds Adjunct Lecturer positions at UNSW Sydney and QUT. He completed his PhD in Telecommunications Engineering from University of South Australia in 2013. He obtained MSc in Electrical and Computer Engineering from Koc University (2008), BSc in Electrical and Electronics Engineering from Bogazici University (2006), and B.A. in Public Administration from Anadolu University (2008).
\end{IEEEbiography}

\vspace{-3\baselineskip}

\begin{IEEEbiography}
    [{\includegraphics[width=1in,height=1.25in,clip,keepaspectratio]{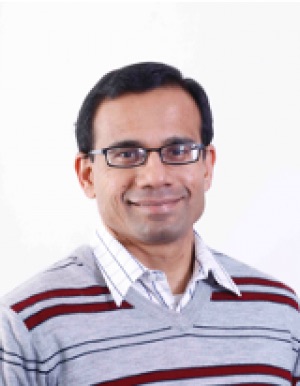}}]{Salil S Kanhere}
    received his M.S. degree and Ph.D. degree from Drexel University in Philadelphia. He is a Professor of Computer Science and Engineering at UNSW Sydney, Australia. He is a Senior Member of the IEEE and ACM, a Humboldt Research Fellow and an ACM Distinguished Speaker. He serves as the Editor in Chief of the Ad Hoc Networks journal and as Associate Editor of IEEE Transactions on Network and Service Management, Computer Communications and Pervasive and Mobile Computing. He has served on the organising committee of many IEEE/ACM international conferences including PerCom, CPS-IOT Week, MobiSys, WoWMoM, MSWiM, and ICBC.
\end{IEEEbiography}

\vspace{-3\baselineskip}

\begin{IEEEbiography}
    [{\includegraphics[width=1in,height=1.25in,clip,keepaspectratio]{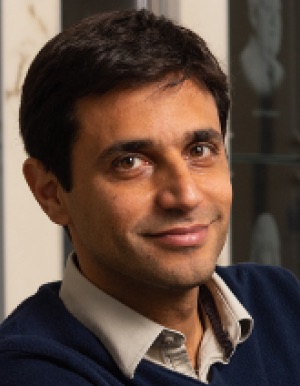}}]{Raja Jurdak}
    is a Professor of Distributed Systems and Chair in Applied Data Sciences at Queensland University of Technology, and Director of the Trusted Networks Lab. He received the PhD in information and computer science from the University of California, Irvine. Prof. Jurdak has published over 190 peer-reviewed publications, including two authored books most recently on blockchain in cyberphysical systems in 2020. He serves on the editorial board of Ad Hoc Networks, and on the organizing and technical program committees of top international conferences, including Percom, ICBC, IPSN, WoWMoM, and ICDCS. He is a conjoint professor with the University of New South Wales, and a senior member of the IEEE.
\end{IEEEbiography}

\vspace{-3\baselineskip}

\begin{IEEEbiography}
    [{\includegraphics[width=1in,height=1.25in,clip,keepaspectratio]{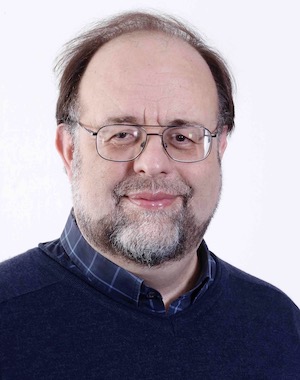}}]{Aleksandar Ignjatovic}
    received the PhD degree in mathematical logic from the University of California, Berkeley. He is an Associate Professor in the School of Computer Science and Engineering at the University of New South Wales, Australia. His current research interests include approximation theory, sampling theory and applied harmonic analysis, algorithm design and applications of mathematical logic to computational complexity theory.
\end{IEEEbiography}





\end{document}